\documentclass{Interspeech2024}
\usepackage{xcolor}
\usepackage{cleveref}
\usepackage{multirow}
\usepackage{bigdelim} 
\usepackage{hyperref}
\usepackage{makecell}
\interspeechcameraready

\newcommand{\astfootnote}[1]{%
  \begingroup
  \renewcommand\thefootnote{\fnsymbol{footnote}}%
  \footnotetext[1]{#1}%
  \endgroup
}

\title{DiveSound: LLM-Assisted Automatic Taxonomy Construction for Diverse Audio Generation}

\name[affiliation={1}]{Baihan}{Li}
\name[affiliation={1}]{Zeyu}{Xie}
\name[affiliation={1}]{Xuenan}{Xu}
\name[affiliation={1}]{Yiwei}{Guo}
\name[affiliation={2}]{Ming}{Yan}
\name[affiliation={2}]{Ji}{Zhang}
\name[affiliation={1}]{Kai}{Yu$^{\ast}$}
\name[affiliation={1}]{Mengyue}{Wu$^{\ast}$}

\address{
  $^1$MoE Key Lab of Artificial Intelligence X-LANCE Lab\\
  Shanghai Jiao Tong University, Shanghai, China\\
  $^2$Institute of Intelligent Computing, Alibaba Group, China}

\email{\{lbh0612, zeyu\_xie, wsntxxn, mengyuewu\}@sjtu.edu.cn}


\begin{document}

\maketitle

% the abstract here must exactly match the abstract entered into the paper submission system
\begin{abstract}
Audio generation has attracted significant attention. Despite remarkable enhancement in audio quality, existing models overlook diversity evaluation. This is partially due to the lack of a systematic sound class diversity framework and a matching dataset. To address these issues, we propose DiveSound, a novel framework for constructing multimodal datasets with in-class diversified taxonomy, assisted by large language models. As both textual and visual information can be utilized to guide diverse generation, DiveSound leverages multimodal contrastive representations in data construction. Our framework is highly autonomous and can be easily scaled up. We provide a text-audio-image aligned diversity dataset whose sound event class tags have an average of 2.42 subcategories. Text-to-audio experiments on the constructed dataset show a substantial increase of diversity with the help of the guidance of visual information. Our samples are available at \href{https://divesounddemo.github.io/}{\textcolor{cyan}{\textit{https://divesounddemo.github.io}}}.

\end{abstract}

\section{Introduction}
\astfootnote{Mengyue Wu and Kai Yu are the corresponding authors.}
% diverse TTA 的一条 motivation: 做 data augmentation 或者 data simulation 的时候, 多样的数据能够让训出来的模型更加鲁棒
% With the rapid advancement of generative models, audio generation has been attracting increasing attention. 
Audio generation, especially text-conditioned audio generation, has been attracting more and more attention in recent years, with numerous applications in film dubbing, game production and virtual reality~\cite{kong2019acoustic}.
% \MY{this is exactly the same with your abstract start, avoid this and you can use shorter sentence in abstract, i.e. combining the current 1st and 2nd sentences in abstract.}
% Text-to-audio (TTA) tasks have diverse applications, such as dubbing for films, background music for games, and virtual reality technologies~\cite{kong2019acoustic}. 
In this scope, the text-to-sound generation focuses on producing sound events beyond speech and music, such as those in the natural world. 
% With the ever growing need, the demand for the diversity of generated audio continues to increase. The increased diversity of the training dataset also contributes to the robustness of trained models. 
Many competitive audio generation methods have been proposed recently, such as DiffSound~\cite{diffsound}, AudioGen~\cite{audiogen}, Make-An-Audio~\cite{makeanaudio} and AudioLDM~\cite{liu2023audioldm}, whose competence in different audio generation tasks has been well demonstrated.
% A popular trend is to utilize pre-trained multi-modal self-supervised learning models via contrastive learning, like CLIP~\cite{radford2021learning} and CLAP~\cite{wu2023large}, to extract representative features from different modalities~\cite{makeanaudio,liu2023audioldm} for generating audio signals.\zy{This statement bears little relevance to the main storyline.} 

\begin{figure}[t]
  \centering
  \includegraphics[width=\linewidth]{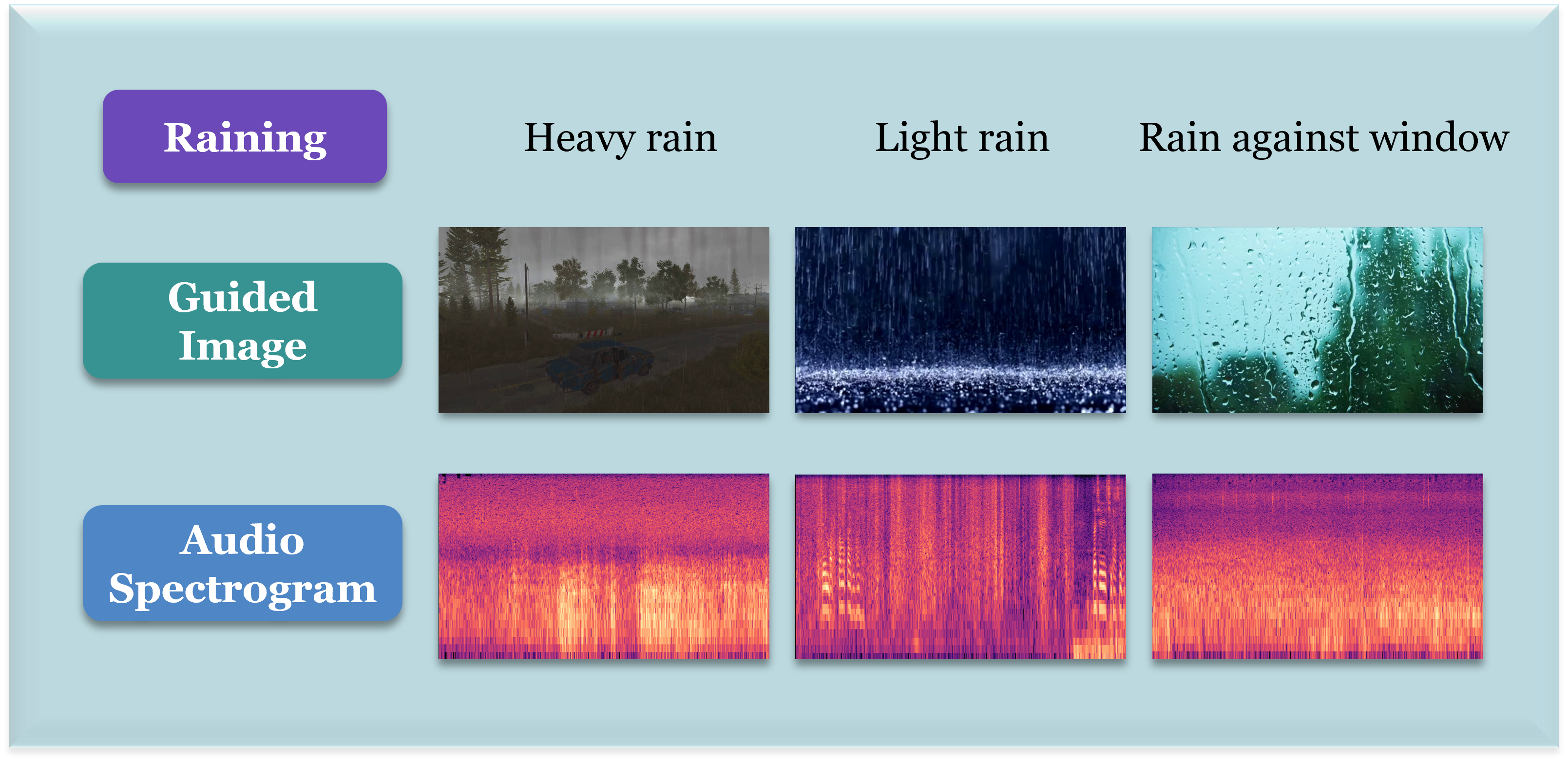}
  \caption{Multimodal text-audio-image dataset samples}
  \label{fig:introsample}
  \vspace{-0.1in}
\end{figure}

However, most prevailing audio generation methods overlook the diversity of the generated sound, and diversity measurement for the same sound event is rarely mentioned. 
Meanwhile, the demand for diversity in audio generation continues to grow with the ever increasing need, because the natural world presents a wide array of sounds, even within the same sound event.
The lack of diversity arises partly due to the long-tailed distribution of existing datasets, which causes models to predominantly learn the most frequent occurrences rather than diverse distributions.
As shown in the \Cref{fig:introsample}, different types of sound may belong to the same sound event, in accordance with different text descriptions or images.
Nevertheless, Existing models can only generate monotonous sounds or a limited range of variation.
%\GYW{The logic here still  seems not very coherent. Ask GPT?} \MY{combine your 1st and 2nd paragraph, it serves 1 purpose: previous works endeavor to increase generated audio quality, xxx, xxx aspect, however they overlooked diversity. give an example and cite your fig.1, i.e. there are different rain sounds while most mainstream models are generating monotonous rain sounds.}

% These methods achieve decent generation specific to different sound events.
% However, in these works, , and the lack of high-quality text-to-audio data pairs is also a significant factor limiting the quality and diversity of sound generation.
% In the work focusing on onomatopoeia generation~\cite{ohnaka2023visual}, images are incorporated as an additional guidance to enhance the generation performance, which inspired our work.
% \MY{this last sentence is not that important, you can delete it and just cite it when saying visual input can work as effective guidance in audio generation. this paragraph should still focus on diversity, dcase did it but limited classes. previous work indicates that involving text/visual input within a class can help enhance diversity. However, currently there's no proper investigation on within class sound diversity. Specifically, there should be a framework that identifies what sound classes exhibit great within diversity. These subclasses should be able to be distinguished textually and visually. another example can be added.}

%  0.这里需要稍微介绍一下 previous work 的做法, 讲清楚缺点, 这个工作应该 reviewer 都没看过, 很容易一头雾水
% \zy{The introduction of visual elements feels rather abrupt. Starting with: `'.}

Some studies have attempted to enhance diversity in the generated audio~\cite{xie2024enhancing}, and they adopt unsupervised algorithms to cluster sound events and integrate visual information to guide generation. This line of research has achieved promising results on small datasets. However, unsupervised algorithms require researchers to explicitly assign labels to each clustered category and find matching images, heavily relying on the subjective interpretation of researchers and thus cannot be scaled up to large datasets. Hence, there is a need for an automated framework to construct diverse datasets, eliminating the necessity for manual intervention and enabling easy scalability.
% The utilization of visual information to guide audio generation has been attempted on small datasets and has yielded promising results. Unsupervised algorithms are used to cluster sound events and results indicates that visual inputs within a class can help enhance generation diversity. However, the unsupervised algorithms requires researchers to explicitly assign labels to each clustered category and find matching images, heavily relying on the subjective interpretation of researchers and thus cannot be scaled up to large datasets. Therefore, a framework that identifies what sound classes exhibit great within diversity is required. 
Due to the inherent alignment between images and audio, better control over audio can be achieved, which has been proved in ~\cite{xie2024enhancing,ohnaka2023visual}. Consequently, the proposed subcategories should be able to be distinguished textually and visually.
% \zy{why it should be? Due to the inherent alignment between images and audio, better control over audio can be achieved.}

% \BH{A 7-category dataset provided by DCASE2023 task7~\cite{choi2023foley} has been used for this approach, which adopts unsupervised algorithms for clustering sound events. Then prototype images for each sound event class are augmented in training and inference process. However, on the one hand, the unsupervised algorithms requires researchers to explicitly assign labels to each clustered category and find matching images, heavily relying on the subjective interpretation of researchers and thus cannot be scaled up to large datasets. On the other hand, the dataset containing only 7 labels is relatively small therefore the model does not have enough practical significance.} \MY{combine this into previous para, introducing the method is important but too much here, 1-2 sentences is enough.}

% 1xnx 为什么突然出现了 VGGSound, 和前文有啥联系?
% \GYW{Too abrupt to introduce VGGSound here.}
% \MY{separate your contribution, you first setup this diversity framework and then a dataset. you can say you use vggsound, but it's not as important as what you did. it's just an audio-visual dataset, one the largest aligned, hence it servers your purpose.}

In this paper, we propose a framework involving three processes: (1) clustering and (2) reasoning to obtain a taxonomy with diverse subcategories, and (3) a matching process to acquire high-quality text-audio-image paired data.  Specifically, the proposed framework \textbf{DiveSound}, aided by large language models (LLMs) to cluster and inference, yields a taxonomy  with diverse subcategories that are distinguishable both visually and textually.
This takes into account the inherent alignment between audio and visual modalities. Leveraging the taxonomy descriptors as intermediaries, we construct an automatic data matching process to establish a dataset.
DiveSound comprises diverse text-audio-image data, where models trained on it demonstrate outstanding diversity while ensuring the quality of audio. In contrast to prior efforts, DiveSound is automatically constructed, facilitating easy scalability without introducing bias from human annotators.
% Our contributions include: proposing a method for automatically constructing audio-text-image datasets that can be scaled up; providing a Taxonomy with finely annotated diverse subcategories of Text-Audio-Image, distinguishable for both text and visual modalities; offering a diverse Text-Audio-Image dataset, DiveSound, whereby models trained on it exhibit good diversity while maintaining audio quality.
% .
The contributions of this work are summarized as follows:
\begin{itemize}
    \item We propose DiveSound, a novel taxonomy to properly define sound diversity and sub-categorize within-class variety, accompanied by an automatic pipeline to align high-quality text-audio-image data. 
    %\item With the proposed taxonomy, we introduce a new multimodal dataset on sound diversity.
    \item Both subjective and objective evaluations demonstrate that by incorporating such a taxonomy-based dataset can enhance generated sound quality and diversity, where visual modality is most helpful in guiding audio generation.
\end{itemize}

\section{Automatic Construction of Diverse Subcategory Taxonomy}
% Dataset Construction\MY{Within Class Audio Diversity, or similar, stress that this is for diversity and then separate the taxnonomy and dataset}}

% The proposed taxonomy construction method follows a two-stage approach: first, we use GPT-4 to cluster and reclassify the sound event labels from VGGSound, and then we utilize CLIP and CLAP models for dataset filtering and integration.

In the construction framework, we initially establish a taxonomy to characterize diverse subcategories for categorizing and organizing high-quality data. Subsequently, we employ an automatic matching process to construct the text-audio-image dataset.
\subsection{Characterizing Sound Diversity Taxonomy with LLM}
% \MY{this section title is not very informative, Diversity taxonomy with LLM, things like this}}
Many sound classes exhibit great within-class diversity however there currently lacks an investigation into what sound classes are mostly diversed and how many sub-categories can be identified.
To cluster and reclassify sound event labels, we employ GPT-4~\cite{achiam2023gpt}, a start-of-the-art LLM with outstanding performance in natural language processing and generation tasks. VGGSound comprises 300 labels, among which certain labels correspond to exceedingly specific sound events for which further diversification is unnecessary. For example, we rarely need to generate the sound of a particular specie of an unusual animal and some special sound events like \textit{`smoke detector beeping'} have no need for classification. At the same time, other labels correspond to sound events emitted by a singular entity and cannot be differentiated through visual stimuli alone (e.g., \textit{`dog barking'} and\textit{ `dog howling'}, which are included in \textit{`dog'}).
% 什么叫 "further diversification is unnecessary"?
% VGGSound comprises 300 labels, among which certain labels correspond to exceedingly specific sound events for which further diversification is unnecessary.
% 这两句为啥是 "conversely" 的关系? 
% 什么是 significant labels, meaningful sound event class?
As a consequence, we need GPT-4 to select the significant labels and regroup them to certain meaningful sound event class, and then generate the subcategories. %The standard for selection of these class is showed in this section.

\subsubsection{Clustering VGGSound labels}

\begin{figure}[htbp]
  \centering
  \includegraphics[width=\linewidth]{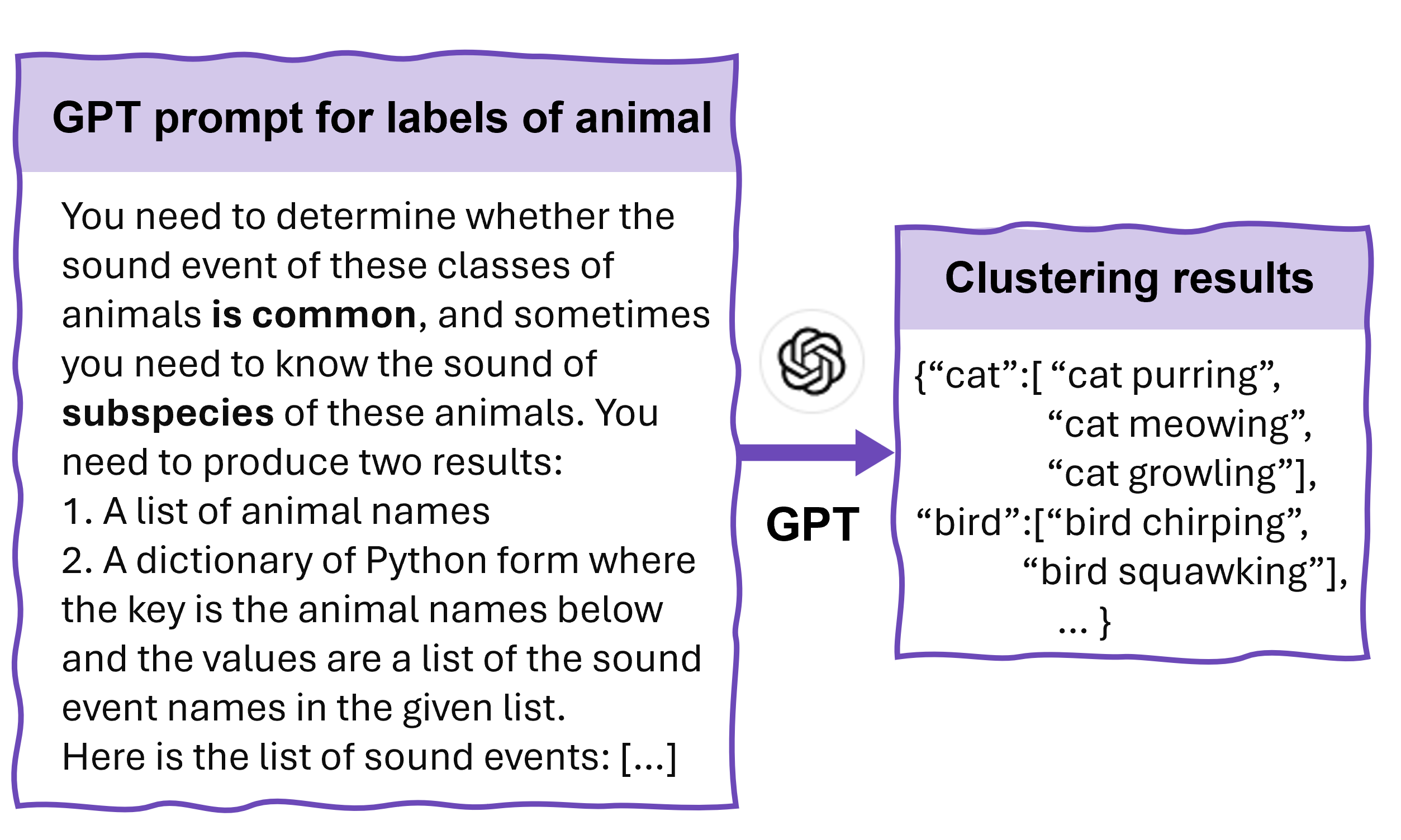}
  \caption{ The process of regrouping the VGGSound labels into new sound event labels, in the example of animals. Sound event labels for other overarching categories are also classified using the same form of prompt.}
  % \MY{i see the two prompts take up much space, you can put them on github link and refer your readers there, if we run out of space in the end}}
  \label{fig:gptprompt1}
\end{figure}

VGGSound~\cite{chen2020vggsound} dataset categorizes auditory labels into nine overarching categories, namely \textit{animals, home, music, nature, people, sports, tools, vehicle} and \textit{others}.
% 什么是 "assemblage of sound events associated within each category"? 什么是 "fidelity of clustering"? 
To enhance the quality of the clustering, this study employed a consistent format of prompt to direct the GPT model to merge different categories separately, which means breaking down a substantial task into smaller components to ensure generation performance.
% Aiming to enhance the quality of the resulted clustering, this study employed a consistent prompt format to direct the GPT model in the assemblage of sound events associated within each distinguished category.
Illustrative representation of the clustering methodology is presented in \Cref{fig:gptprompt1}.
% 什么是 "VGGSound compendium"?
After clustering, a refined seletion with 69 \textbf{new classes}, containing 83 distinct labels from the VGGSound.
% After clustering, a refined selection encompassing 83 distinct labels from the VGGSound labels was conducted.
% % xnx 怎么 reallocate 的?
% These specifically chosen labels were then reallocated, culminating in the formation of 69 new sound event classifications.

\subsubsection{ Reasoning subcategories of each class}

\begin{figure}[h]
  \centering
  \includegraphics[width=\linewidth]{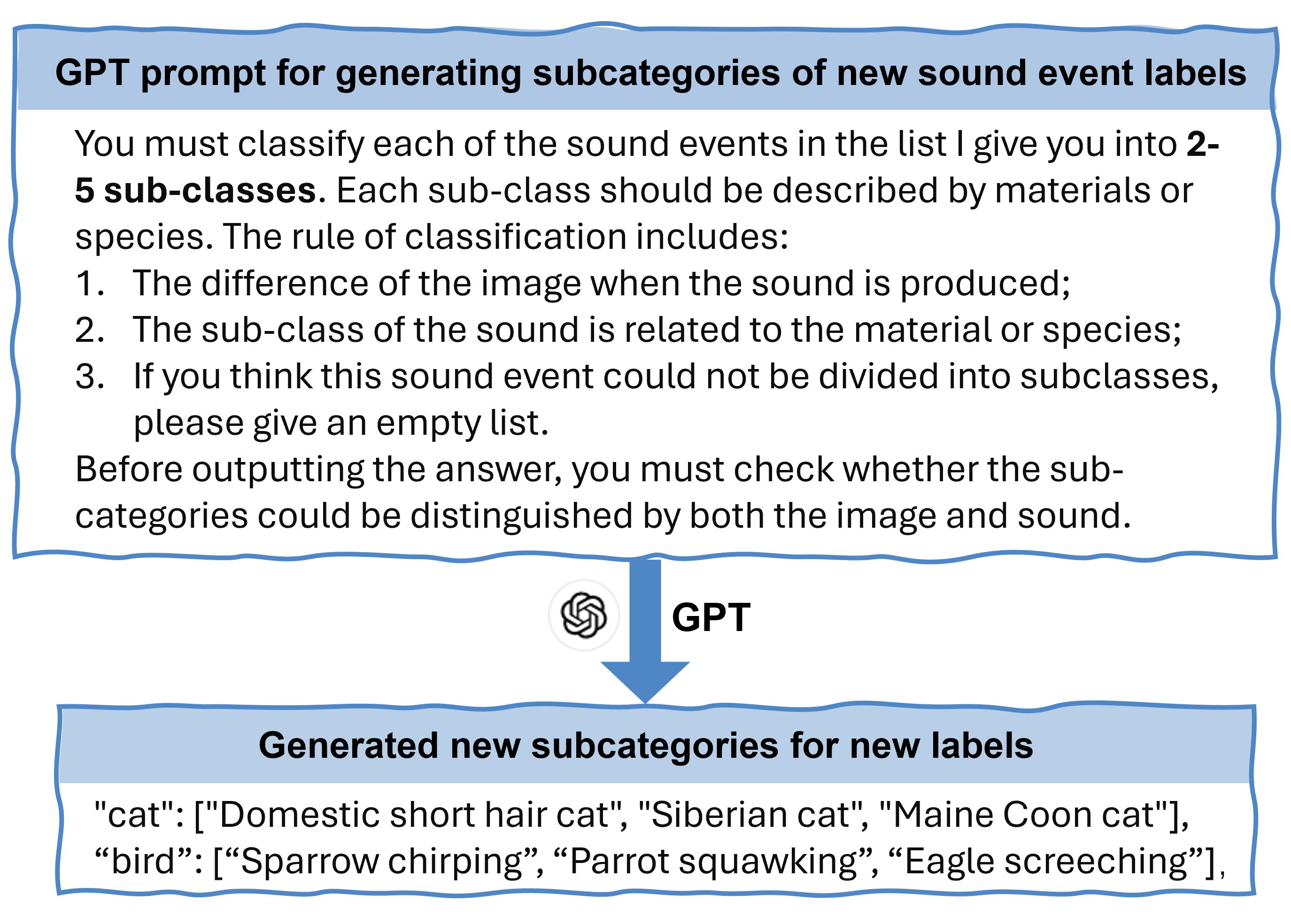}
  \caption{The prompt and examples of editing classifications and creating new subcategories. Three important rules are listed.}
   % \MY{to many text in this fig, only showing the prominent ones, the font size should be equal or only 1 size smaller than the main text. use color blocks to highlight your text.}
  \label{fig:gptprompt2}
\end{figure}

After obtaining the selected and summarized labels, we aim for GPT to reclassify these categories and generate more diverse labels.
% reclassify and generate 为什么是 filtering?
During this reclassification and generation process, we aim to leverage the visual information included in VGGSound.
Therefore, we have incorporated into the classification criteria that each subcategory within each major class should be distinguishable both visually and auditorily. The specific form of the prompt and classification results is demonstrated in \Cref{fig:gptprompt2}.
% 这里的 refinement 和 reclassify & generate / filter 是一回事吗?
After this selection and generation work by GPT, we ultimately obtained 40 valid new sound event classes.

\subsection{Automatic Text-Audio-Image Data Matching Process}\label{subcategory}
\begin{figure*}[htbp]
  \centering
  \centerline{\includegraphics[width=0.9\textwidth]{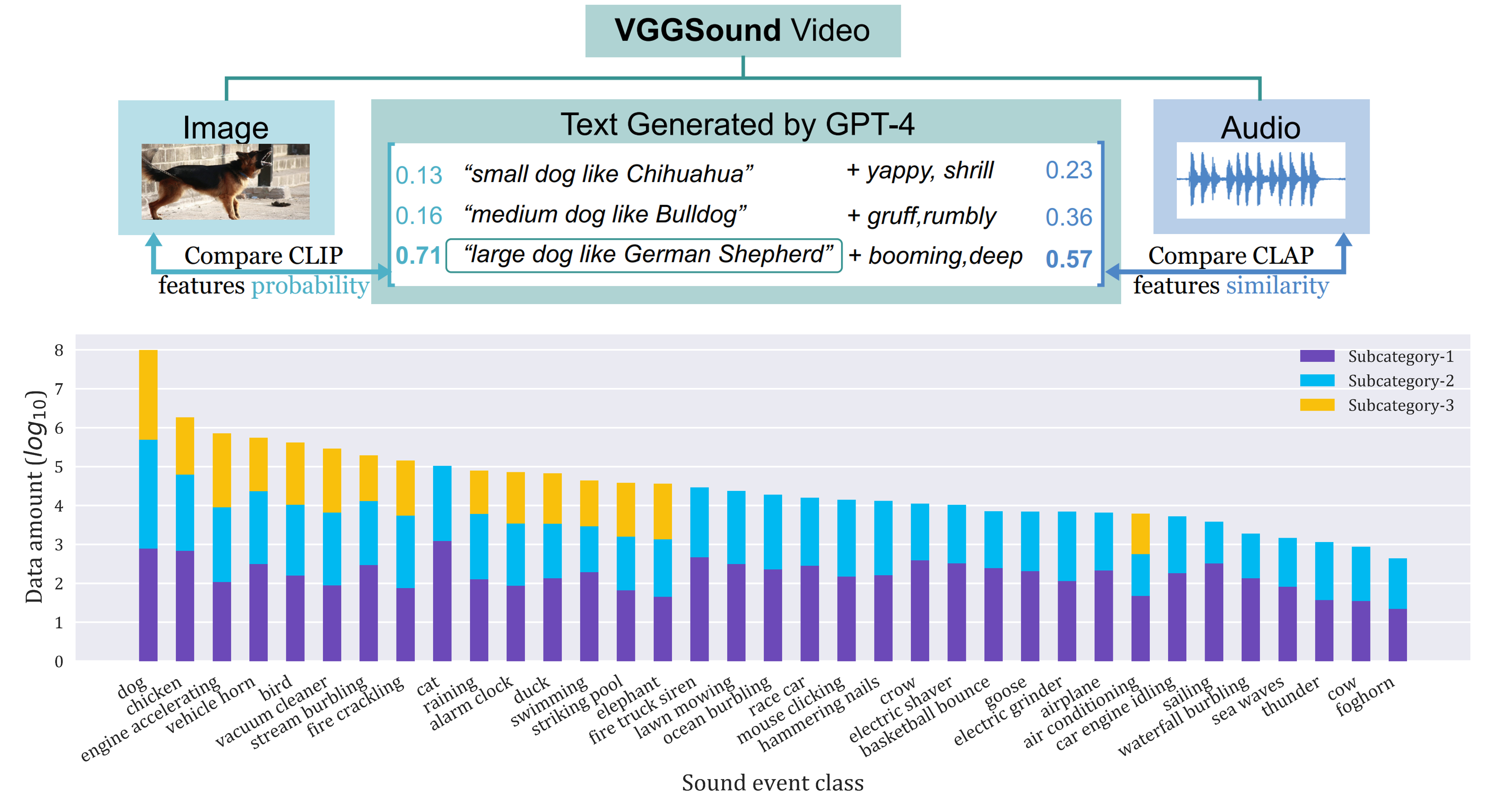}}
  \caption{ Top: the automated matching process of text-audio data pairs. The example here uses the new class \textit{dog} to demonstrate how an audio clip is matched with its corresponding text data pair. Bottom: the statistics of the newly selected dataset, including the 35 class labels with an average of 2.42 subcategories.}
  %\MY{color contrast is not strong enough, teal and blue can easily be confused}
  \label{fig:dataset}
\end{figure*}

% \begin{figure*}[htbp]
%   \centering
%   \centerline{\includegraphics[width=\textwidth]{figures/stacked_bar_plot.png}}
%   \caption{ Bar chart illustrates the statistics of the newly selected dataset, including the 35 class labels with an average of $2.42$ subcategories.}
%   %\MY{color contrast is not strong enough, teal and blue can easily be confused}
%   \label{fig:dataset}
% \end{figure*}
To construct text-audio-image data pair, we use text descriptions serving as a bridge for connecting visual and audio modalities, thus CLIP~\cite{radford2021learning} and CLAP~\cite{wu2023large} are utilized to create a shared space for text-image and text-audio respectively. 
% 什么 process? 
% gyw: 什么 process? 
% processssss!!!!!!
The pipeline of this auto-matching process is illustrated in \Cref{fig:dataset}, which could be summarized by 3 steps.

% video 是怎么 select 的?
% video 和 image
\noindent\textbf{Step 1: Connect text and image features}
%\MY{Remember always say why, don't start your paragraph from other people's work: rephrase 
We use text as an intermediate pivot to bridge visual and auditory information. In step 1, we focus on matching text with corresponding images.
%As VGGSound is a dataset of video clips, 
After obtaining the reclassification result, we utilized CLIP to extract features from randomly selected frames.
Simultaneously, we extracted features from the subcategories' text of each new class generated by GPT.
This serves two purposes: 
\begin{itemize}
    \item Calculate the probability of images being classified as various texts, identifying the potential sound events depicted in each audio's visual frame.
    \item Find the image with the highest similarity to each class of text and image, serving as representative image information for that audio class.
\end{itemize}

% 和 Figure 4 要有联系, 比如 Figure 4 中哪部分是 subclass, 哪部分是 adj
\noindent\textbf{Step 2: Connect text and audio features}
% \MY{
A similar approach is utilized to align text with audio features. 
We empoly CLAP to identify the closest subclass within each new class for every audio.
To enhance classification accuracy, we incorporated 2-4 adjectives generated by GPT-4 for each subclass during similarity calculations. The Top diagram in \Cref{fig:dataset} shows that the text description of each subcategory on the left is concatenated with the generated adjectives on the right.  

By computing the similarity between audio and text data, we determine the subclass text that best matches each audio within the respective new class.

\noindent\textbf{Step 3: Match the text-audio-image data}
After obtaining text descriptions with the highest similarity scores to the audio and its associated visual frame respectively, we proceed with audio matching and filtering.
When both image and audio features are most similar to the same text description, we establish a text-audio data pair, assigning audio data to this subcategory.
The new class should have enough audio data, as a result if the number of audio data corresponding to a subclass is less than 20, we discard that subclass.

\noindent\textbf{End Product} This process resulted in 35 classes, with an average of $2.42$ subcategories per class and over 10K clips.
The details of each class and the class labels are shown in \Cref{fig:dataset}. These subcategories are not only audibly diversed from each other, but also visually distinguishable and can be described with varying text. For instance the different species of dogs could be distinguished both by their appearance and their barking sound. Hereby, we identify sound classes that exhibit distinguishable inner-diversity and provide corresponding textual descriptions as well as visual supplements.

\section{ Diverse Audio Generation Model}
% Multi-Modality-Guided Audio Generation\MY{diversity! Diversed Audio Generation}}
% 稍微概括一下这个 Section 要介绍模型了
In this section, we introduce the generation model, which is similar to \textit{Xie et al.}~\cite{xie2024enhancing} where latent diffusion model is used for generation.
Text and visual features are embedded into training with a fusion module as diversity guidance.
% \MY{our generation framework is similar to xxxx, general framework introduction and then say that since we have text and visual features as input, we have a modality fusion module, followed by a latent diffision model domonating the generation process.}
\vspace{-0.1in}
\subsection{Modality Fusion}
\vspace{-0.1in}
The information from different modalities is fused into the model in the training process.
Initially, the new class labels $\mathcal{C}_i$ are encoded into embedding $\mathcal{E}_\text{label}$ using a lookup table.
Then, the audio data from each subcategory $\mathcal{C}_{i,j}$ of each class obtained in \Cref{subcategory} is augmented with the descriptive text $v_{\text{text}}$ and representative image $v_{\text{image}}$ features extracted by CLAP and CLIP respectively.
% \zy{The formula can be omitted; simply stating "concatenate with class information" suffices.}
% xnx 公式里 v_{text}, v_{image} 是啥? 公式得到的 \mathcal{E} 干啥用的?
The embedding of fusion inputs concatenated with visual and textual information are as follows:
% \begin{equation}
%     \begin{split}
%         \mathcal{E}_{baseline} = \mathcal{E}_{label}\\
%         \mathcal{E}_{text} = \mbox{concat}(\mathcal{E}_{label}, {\mbox{CLAP}(v_{text})})\\ 
%         \mathcal{E}_{image} = \mbox{concat}(\mathcal{E}_{label}, {\mbox{CLIP}(v_{image})})
%     \end{split}
% \end{equation}
\[
\begin{aligned}
    \mathcal{E}_\text{base} &= \mathcal{E}_\text{label}\\
    \mathcal{E}_\text{text} &= \mbox{concat}(\mathcal{E}_\text{label}, {\mbox{CLAP}(v_\text{text})})\\ 
    \mathcal{E}_\text{image} &= \mbox{concat}(\mathcal{E}_\text{label}, {\mbox{CLIP}(v_\text{image})})
\end{aligned}
\]
% xnx 如果要简略介绍 LDM 的话, 没必要写这么多, 输入输出是啥, 尤其是和前面定义的 \mathcal{E} 有啥关联写清楚
\begin{table*}[htbp]
\centering
\caption{Objective and subjective evaluation metrics, tested on baseline, text-guided and image-guided systems. Real audios are considered as ground truth. \textit{p}-value for the t-trest comparing MSD results of the image-guided system with baseline and text-guided systems are $0.0055$ and $0.0032$ respectively. * indicates significance $<0.05$   and ** for $<0.005$  }
% \label{evaltable} 
% \renewcommand{\arraystretch}{1.1}
% \begin{tabular}{c|cccc|ccc}
    
%     \toprule
%     \textbf{System}& \multicolumn{4}{c|}{\textbf{Objective}}& \multicolumn{3}{c}{\textbf{Subjective}} \\ 
%     \cmidrule(lr){2-5}\cmidrule(lr){6-8}
%     &Quality (FAD) $\downarrow$ & & Diversity (MSD) \uparrow &  & Accuracy $\uparrow$ &Diversity $\uparrow$ &Naturalness $\uparrow$\\

%    \midrule
%    %\multicolumn{6}{c}{\textbf{VAE+Diffusion}} \\
%    %\midrule
%     Ground truth &- &\multirow{3}{*}{\rdelim[{3}{*}[*]} &-& & $4.00$   & $4.07$   & $3.87$ \\
%    Baseline &  $15.20\pm0.93$& $10.92\pm0.17$* &   $3.59$           &  $3.06$           & $3.43$\\ 

%    Text-guided &   $14.60\pm1.02$ &$11.30\pm0.21$** & $\mathbf{3.69}$           & $3.40$   & $3.46$\\
%    Image-guided &   $\mathbf{13.13\pm0.79}$ & & $\mathbf{12.62\pm0.13}$ & $\mathbf{3.69}$     &  $\mathbf{3.77}$   &$\mathbf{3.53}$ \\

%     \bottomrule
%     \end{tabular}
\label{evaltable}  
\renewcommand{\arraystretch}{1.1}
\begin{tabular}{c|cc|ccc}
    
    \toprule
    \textbf{System}& \multicolumn{2}{c|}{\textbf{Objective}}& \multicolumn{3}{c}{\textbf{Subjective}} \\ 
    \cmidrule(lr){2-3}\cmidrule(lr){4-6}
    &Quality (FAD) $\downarrow$ & Diversity (MSD) $\uparrow$ &Accuracy $\uparrow$ &Diversity $\uparrow$ &Naturalness $\uparrow$\\

   \midrule
   %\multicolumn{6}{c}{\textbf{VAE+Diffusion}} \\
   %\midrule
    Ground truth &- &- & $4.00$   & $4.07$   & $3.87$ \\
   Baseline  &  $15.20\pm0.93$&  $10.92\pm0.17$* &     $3.59$           &  $3.06$           & $3.43$\\ 
   Text-guided &    $14.60\pm1.02$ &$11.30\pm0.21$** & $\mathbf{3.69}$           & $3.40$   & $3.46$\\
   Image-guided &   $\mathbf{13.13\pm0.79}$ &  $\mathbf{12.62\pm0.13}$  & $\mathbf{3.69}$     &  $\mathbf{3.77}$   &$\mathbf{3.53}$ \\
    \bottomrule
    \end{tabular}
\end{table*}
\vspace{-0.2in}
\subsection{Latent Diffusion Model}

We utilized the Latent Diffusion Model (LDM) as the backbone structure, comprising Variational Autoencoder (VAE), diffusion, and vocoder models, which have demonstrated excellent performance in audio generation tasks~\cite{audioldm2,liu2023audioldm}.
The VAE model is employed to extract representations of the audio, aiming to reduce computational complexity. It comprises an encoder, which compresses the Mel spectrogram into the latent space, and a decoder, which reconstructs the spectrogram based on samples. 
Finally, the waveform is reconstructed by the vocoder.
A diffusion model is employed to predict the latent representation, relying on conditional inputs $\mathcal{E}_\text{base}/ \mathcal{E}_\text{text} / \mathcal{E}_\text{image}$.
In detail, a forward process continuously introduces noise into the latent representation, resulting in Gaussian noise. 
A reverse process gradually eliminating noise from the Gaussian noise.

We utilize the VAE introduced by AudioLDM~\cite{liu2023audioldm}.
Our LDM shares a similar architecture to that of Ghosal D et al.~\cite{ghosal2023text} but with a reduced parameter count, employing attention dimensions of $\{4, 8, 16, 16\}$ and block channels of $\{128, 256, 512, 512\}$. The LDM is trained using the AdamW optimizer for $80$ epochs with a linear decay scheduler. We set the learning rate to $3\times10^{-5}$ and the classifier free guidance scale to 3.

\
% section{Experiments}
% \subsection{Generation Framework}\MY{this can actually go up into generation section}
% \subsection{Experimental Setup}
% We utilize the VAE introduced by AudioLDM~\cite{liu2023audioldm}.
% The Latent Disentanglement Model (LDM) shares a similar architecture to that of Ghosal D et al.~\cite{ghosal2023text} but with a reduced parameter count, employing attention dimensions of $\{4, 8, 16, 16\}$ and block channels of $\{128, 256, 512, 512\}$. The LDM is trained using the AdamW optimizer for $80$ epochs, with a learning rate set to $3\times10^{-5}$ and a linear decay scheduler. Classifier-free guidance~\cite{ho2022classifier,nichol2021glide} is adopted for controllable generation. The guidance scale is set to 3.
\vspace{-0.2in}
\subsection{Evaluation Metrics}

We use both objective and subjective metrics to assess the quality and diversity of the generated audio, thus comparing the impact of different modalities on audio generation. 

\textbf{Objective-Quality}
Fr$\acute{\text{e}}$chet Audio Distance (FAD)~\cite{kilgour2019frechet} is an objective metric widely used in the field of text audio generation to measure the quality of generated audio. It is applied to measure the model performance in terms of generation quality by measuring the distance between generated audio and real audio. We chose it as our metric for objective-quality evaluation to measure model performance by calculating the average value of the model over 35 classes of generated audio.

\textbf{Objective-Diversity}
Having ensured that the quality of the generated audio clips are comparable, Mean Squared Distance (MSD) is used as a measure of the diversity of the generated audio~\cite{ohnaka2023visual}. To calculate MSD between generated audio clips, the self-supervised BEATs~\cite{chen2022beats} model is used to extract audio features that are not specific to categories.  After the features have been extracted, we calculate the distance between any two audio clips in each class and use the average distance to represent an objective measure of the diversity of the audio in this class.

\textbf{Subjective-Evaluation}
The Mean Opinion Score (MOS) is evaluated based on three criteria: event accuracy, generation diversity, and audio naturalness. We randomly selected 20\% of the classes for subjective evaluation. 10 normally-hearing participants aged 22-24 who are familiar with audio/speech subjective evaluation tasks are selected as evaluators. %They have been trained in subjective evaluation experiments across various tasks and 
They were asked to listen to concatenated audio segments, each 30 seconds long, consisting of generated audio or real audio inferred by the same system. 
% \MY{and how they match their score 1-5 to diversity extent?}. 
This method is inspired by the generated image diversity evaluation~\cite{ramesh2022hierarchical} and allows evaluators to compare the diversity between audios generated by the same system. They then scored the entire audio segment with a marking scheme from 1 to 5. In terms of the diversity, they need to compare the difference between the . The final score was calculated as the average of the ratings from 10 evaluators.

\subsection{Results}
% The results are presented in \Cref{evaltable} which elaborates the FAD and MSD values for the baseline, text-guided and audio-guided system and presents the overall subjective MOS diversity. 
%We repeated 3 times experiments for baseline, text-guided and audio-guided systems. 
We compared vanilla system (baseline) results with text- and audio-guided generation models, with all experiments repeated using 3 random seeds.
We calculate the objective metrics for all the systems. In \Cref{evaltable}, we present the average and standard deviation of FAD and MSD and the results of the MOS evaluation.

\noindent\textbf{Diversity Enhancement}
Both objective and subjective metrics in \Cref{evaltable} shows an enhancement in the diversity of generated audios with the guidance of both textual and visual information.  

In the MSD evaluation, it is observed that the diversity of text-guided system is slightly higher than that of the baseline system, while the diversity metric of the image-guided system is notably elevated the previous two systems. In subjective evaluations, despite the diversity scores of the generated audio being lower than those of real audio, the image-guided system still outperforms the other two approaches. Low standard deviations show that the image-guided system has the strongest stability. The \textit{p}-values 
% for the t-test comparing the MSD results of the image-guided system with the baseline and text-guided systems are 0.0055 and 0.0032, respectively, which 
indicates the guidance of image has a significant impact on the enhancement of the diversity of audio generation.

The subjective results indicate that the in-class diversity of the proposed dataset is guaranteed, which means that our framework is able to construct a high-quality dataset. According to both objective and subjective metrics, the improvement of generation diversity from baseline to systems with different modalities shows the effectiveness of training with model fusion. Furthermore, the visual information have a better performance than the textual information in terms of guiding sound generation.

% \begin{table}[htbp]
%     \centering
%     \caption{Objective evaluation}
%     \label{table1}
%     \begin{tabular}{c|cc}
    
%     \toprule
%     Model & Quality (FAD) $\downarrow$ & Diversity (MSD) $\uparrow$\\

%    \midrule
%    %\multicolumn{6}{c}{\textbf{VAE+Diffusion}} \\
%    %\midrule
   
%    Baseline &  $15.90$& $11.52$  \\
%    Text-guided &   $14.29$ &$11.67$ \\
%    Image-guided &   $\mathbf{13.69}$ & $\mathbf{12.77}$ \\
%     \bottomrule
%     \end{tabular}
% \end{table}

\noindent\textbf{Overall Quality}
In \Cref{evaltable}, FAD and subjective scores in terms of accuracy and naturalness indicate that incorporating image or text information has a positive effect on enhancing the quality of generated audio. Furthermore, the image-guided system performs the best, followed by the text-guided system, both slightly outperforming the baseline.

According to the objective metrics, text-guided and image-guided systems also show their better controllability of generated audio. That's because while the textual information is augmented in the embedding and provides more information for each subcategory. Furthermore, image contains more aligned information to audio than text, leading to lower FAD between ground truth and image-guided system.  
% \MY{prepare an anonymous github link and put it in abstract or intro, you can later (after ddl) to add content on the link, include some sound clips and its corresponding tag and image. if we have space, better to include a figure here, showing that AudioLDM/other models generate 1 kind of sound (one spectrogram), with subcategory text labels, spectrograms; with visual input, spectrograms}
% \begin{table}[t]
% \centering
% \caption{Subjective evaluation, tested on the incorporating visual information.}
% \label{table2}
%     \begin{tabular}{c|ccc}
%     \toprule
%      System  & Accuracy $\uparrow$ &Diversity $\uparrow$ &Naturalness $\uparrow$\\
%     \midrule
%    Ground truth &  $4.00$   & $4.07$   & $3.87$ \\
%    Baseline &   $3.59$           &  $3.06$           & $3.43$\\ 
%    Text-guided & $\mathbf{3.69}$           & $3.40$   & $3.46$\\
%    Image-guided & $\mathbf{3.69}$     &  $\mathbf{3.77}$   &$\mathbf{3.53}$ \\
%    \bottomrule
%     \end{tabular}
% \end{table}
\vspace{-0.1in}
\section{Conclusion}

The aim of this study is to propose a method for automatically matching text-audio-image data pairs and creating multimodal datasets, with the further objective of enhancing the diversity of text-to-audio generation. We introduce the DiveSound framework, composed of a 3-stage approach using a large language model and pretrained CLIP and CLAP models. Based on the reclassification of VGGSound, a dataset comprising multiple categories including subcategories and multimodal information, is introduced. Furthermore, by training a diffusion-based text-to-audio model on the proposed dataset, we demonstrate the positive impact of incorporating different modality information on the quality and diversity of generated audio and image-guided generation outperforms other methods. However, the quality of subcategories' text description is constrained by the knowledge of LLM and the accuracy of data matching is also limited by the representation ability of CLIP and CLAP. In our future study, we will apply our DiveSound framework on more data and explore different method for augmenting multimodal information, in order to furtherly increase the sound generation diversity.
% \GYW{Lack of limitation and future work. Limitation is required by interspeech.}
% \ifinterspeechfinal
%      The Interspeech 2024 organisers
% \else
%      The authors
% \fi
% would like to thank ISCA and the organising committees of past Interspeech conferences for their help and for kindly providing the previous version of this template.
\section{Acknowledgements}
This work was supported by National Natural Science Foundation of China (Grant No.92048205), the Key Research and Development Program of Jiangsu Province~(No.BE2022059), Guangxi major science and technology project~(No. AA23062062) and Alibaba Innovative Research. 

\bibliographystyle{IEEEtran}
\bibliography{mybib}

% Generated by IEEEtran.bst, version: 1.13 (2008/09/30)
\begin{thebibliography}{10}
\providecommand{\url}[1]{#1}
\csname url@samestyle\endcsname
\providecommand{\newblock}{\relax}
\providecommand{\bibinfo}[2]{#2}
\providecommand{\BIBentrySTDinterwordspacing}{\spaceskip=0pt\relax}
\providecommand{\BIBentryALTinterwordstretchfactor}{4}
\providecommand{\BIBentryALTinterwordspacing}{\spaceskip=\fontdimen2\font plus
\BIBentryALTinterwordstretchfactor\fontdimen3\font minus \fontdimen4\font\relax}
\providecommand{\BIBforeignlanguage}[2]{{%
\expandafter\ifx\csname l@#1\endcsname\relax
\typeout{** WARNING: IEEEtran.bst: No hyphenation pattern has been}%
\typeout{** loaded for the language `#1'. Using the pattern for}%
\typeout{** the default language instead.}%
\else
\language=\csname l@#1\endcsname
\fi
#2}}
\providecommand{\BIBdecl}{\relax}
\BIBdecl

\bibitem{kong2019acoustic}
Q.~Kong, Y.~Xu, T.~Iqbal, Y.~Cao, W.~Wang, and M.~D. Plumbley, ``Acoustic scene generation with conditional samplernn,'' in \emph{ICASSP 2019 - 2019 IEEE International Conference on Acoustics, Speech and Signal Processing (ICASSP)}, 2019, pp. 925--929.

\bibitem{diffsound}
D.~Yang, J.~Yu, H.~Wang, W.~Wang, C.~Weng, Y.~Zou, and D.~Yu, ``{DiffSound}: Discrete diffusion model for text-to-sound generation,'' \emph{{IEEE} {ACM} Trans. Audio Speech Lang. Process.}, vol.~31, pp. 1720--1733, 2023.

\bibitem{audiogen}
F.~Kreuk, G.~Synnaeve, A.~Polyak, U.~Singer, A.~D{\'{e}}fossez, J.~Copet, D.~Parikh, Y.~Taigman, and Y.~Adi, ``{AudioGen}: Textually guided audio generation,'' in \emph{The Eleventh International Conference on Learning Representations, {ICLR} 2023, Kigali, Rwanda, May 1-5, 2023}, 2023.

\bibitem{makeanaudio}
R.~Huang, J.~Huang, D.~Yang, Y.~Ren, L.~Liu, M.~Li, Z.~Ye, J.~Liu, X.~Yin, and Z.~Zhao, ``{Make-An-Audio}: Text-to-audio generation with prompt-enhanced diffusion models,'' in \emph{International Conference on Machine Learning, {ICML} 2023, 23-29 July 2023, Honolulu, Hawaii, {USA}}, ser. Proceedings of Machine Learning Research, vol. 202.\hskip 1em plus 0.5em minus 0.4em\relax {PMLR}, 2023, pp. 13\,916--13\,932.

\bibitem{liu2023audioldm}
H.~Liu, Z.~Chen, Y.~Yuan, X.~Mei, X.~Liu, D.~P. Mandic, W.~Wang, and M.~D. Plumbley, ``{AudioLDM}: Text-to-audio generation with latent diffusion models,'' in \emph{International Conference on Machine Learning, {ICML} 2023, 23-29 July 2023, Honolulu, Hawaii, {USA}}, ser. Proceedings of Machine Learning Research, vol. 202.\hskip 1em plus 0.5em minus 0.4em\relax {PMLR}, 2023, pp. 21\,450--21\,474.

\bibitem{xie2024enhancing}
Z.~Xie, B.~Li, X.~Xu, M.~Wu, and K.~Yu, ``Enhancing audio generation diversity with visual information,'' \emph{arXiv preprint arXiv:2403.01278}, 2024.

\bibitem{ohnaka2023visual}
H.~Ohnaka, S.~Takamichi, K.~Imoto, Y.~Okamoto, K.~Fujii, and H.~Saruwatari, ``Visual onoma-to-wave: environmental sound synthesis from visual onomatopoeias and sound-source images,'' in \emph{ICASSP 2023-2023 IEEE International Conference on Acoustics, Speech and Signal Processing (ICASSP)}.\hskip 1em plus 0.5em minus 0.4em\relax IEEE, 2023, pp. 1--5.

\bibitem{achiam2023gpt}
J.~Achiam, S.~Adler, S.~Agarwal, L.~Ahmad, I.~Akkaya, F.~L. Aleman, D.~Almeida, J.~Altenschmidt, S.~Altman, S.~Anadkat \emph{et~al.}, ``{GPT}-4 technical report,'' \emph{arXiv preprint arXiv:2303.08774}, 2023.

\bibitem{chen2020vggsound}
H.~Chen, W.~Xie, A.~Vedaldi, and A.~Zisserman, ``{VGGSound}: A large-scale audio-visual dataset,'' in \emph{ICASSP 2020-2020 IEEE International Conference on Acoustics, Speech and Signal Processing (ICASSP)}.\hskip 1em plus 0.5em minus 0.4em\relax IEEE, 2020, pp. 721--725.

\bibitem{radford2021learning}
A.~Radford, J.~W. Kim, C.~Hallacy, A.~Ramesh, G.~Goh, S.~Agarwal, G.~Sastry, A.~Askell, P.~Mishkin, J.~Clark \emph{et~al.}, ``Learning transferable visual models from natural language supervision,'' in \emph{International conference on machine learning}.\hskip 1em plus 0.5em minus 0.4em\relax PMLR, 2021, pp. 8748--8763.

\bibitem{wu2023large}
Y.~Wu, K.~Chen, T.~Zhang, Y.~Hui, T.~Berg-Kirkpatrick, and S.~Dubnov, ``Large-scale contrastive language-audio pretraining with feature fusion and keyword-to-caption augmentation,'' in \emph{ICASSP 2023-2023 IEEE International Conference on Acoustics, Speech and Signal Processing (ICASSP)}.\hskip 1em plus 0.5em minus 0.4em\relax IEEE, 2023, pp. 1--5.

\bibitem{audioldm2}
H.~Liu, Q.~Tian, Y.~Yuan, X.~Liu, X.~Mei, Q.~Kong, Y.~Wang, W.~Wang, Y.~Wang, and M.~D. Plumbley, ``{AudioLDM} 2: Learning holistic audio generation with self-supervised pretraining,'' \emph{CoRR}, vol. abs/2308.05734, 2023.

\bibitem{ghosal2023text}
D.~Ghosal, N.~Majumder, A.~Mehrish, and S.~Poria, ``Text-to-audio generation using instruction-tuned {LLM} and latent diffusion model,'' \emph{arXiv preprint arXiv:2304.13731}, 2023.

\bibitem{kilgour2019frechet}
K.~Kilgour, M.~Zuluaga, D.~Roblek, and M.~Sharifi, ``Fr{\'e}chet audio distance: A reference-free metric for evaluating music enhancement algorithms.'' in \emph{INTERSPEECH}, 2019, pp. 2350--2354.

\bibitem{chen2022beats}
S.~Chen, Y.~Wu, C.~Wang, S.~Liu, D.~Tompkins, Z.~Chen, W.~Che, X.~Yu, and F.~Wei, ``{BEATs}: Audio pre-training with acoustic tokenizers,'' in \emph{International Conference on Machine Learning, {ICML} 2023, 23-29 July 2023, Honolulu, Hawaii, {USA}}, ser. Proceedings of Machine Learning Research, vol. 202.\hskip 1em plus 0.5em minus 0.4em\relax {PMLR}, 2023, pp. 5178--5193.

\bibitem{ramesh2022hierarchical}
A.~Ramesh, P.~Dhariwal, A.~Nichol, C.~Chu, and M.~Chen, ``Hierarchical text-conditional image generation with clip latents,'' \emph{arXiv preprint arXiv:2204.06125}, vol.~1, no.~2, p.~3, 2022.

\end{thebibliography}

\end{document}